\documentclass[12pt,fleqn]{article}
\usepackage{cite}
\usepackage{amssymb,epsfig,graphicx}
\usepackage{times}
\def\openone{\leavevmode\hbox{\small1\kern-3.6pt\normalsize1}}
\def \text{\mathrm}

\def \be{\begin{equation}}
\def \ee{\end{equation}}
\def \bea{\begin{eqnarray}}
\def \eea{\end{eqnarray}}
\def \bit{\begin{itemize}}
\def \eit{\end{itemize}}
\def \baR{\begin{array}}
\def \eaR{\end{array}}

\def \B{\bar{B}}

\def \Bd{\bar{B}_d \to \mu^+\mu^-}
\def \Bs{\bar{B}_s \to \mu^+\mu^-}
\def \Bq{\bar{B}_q \to  l^+ l^-}
\def \Bds{\bar{B}_{d,s} \to  \mu^+ \mu^-}
\def \br{{\cal B}\,}
\def \f{\frac}
\def \fig#1{Fig.~\ref{#1}}

\def \ml#1{M^2_{\tilde{#1}_{L}}}
\def \M{{\cal M}}

\def \op{{\cal O}}
\def \rf{Ref.~\cite}
\def \rfs{Refs.~\cite}
\def \sec#1{Sec.~\ref{#1}}
\def \vckm{V_{\text{CKM}}}
\def \chargino{\tilde{\chi}^{\pm}}
\def \cl#1{{#1\%\ \mathrm{C.L.}}}

\def \diag{{\mathrm{diag}}}

\def \ea{{\it et al.}}

\def \eq#1{Eq.~(\ref{#1})}

\def \fig#1{Fig.~\ref{#1}}

\def \Oi{{\mathcal O}}

\def \sm{\mathrm{SM}}
\def \rf{Ref.~\cite}
\def \rfs{Refs.~\cite}
\def \sec#1{Sec.~\ref{#1}}

\def\euro#1#2#3{{Eur. Phys. J. C} {\bf #1}, #3 (#2)}

\def\np#1#2#3{{Nucl.~Phys.}~{\bf B#1}, #3 (#2)}

\def\pl#1#2#3{{Phys.~Lett. B}~{\bf #1}, #3 (#2)}

\def\prd#1#2#3{{Phys.~Rev. D}~{\bf #1}, #3 (#2)}

\begin{document}

\mathindent40pt

\begin{flushright}
TUM-HEP-487/02\\
PSI-PR-02-14
\end{flushright}
\vskip 0.2in
\begin{center}
\setlength{\baselineskip}{0.35in} 
{\boldmath\bf \Large Neutral Higgs boson contributions to the decays $\Bds$ 
       in the MSSM  at large $\tan\beta$}\\[1cm]
\setlength {\baselineskip}{0.2in}
{\bf J\"org Urban\footnote{To appear in the proceedings of the
10th International Conference on Supersymmetry and Unification of Fundamental
Interactions (SUSY02). Talk based on collaborations 
with C. Bobeth, T. Ewerth and F. Kr\"uger \cite{{beku02},{cscp}}.}}\\[1mm]
\emph{{Technische Universit\"at M\"unchen,D-85748  Garching, Germany}\\
      {Paul Scherrer Institut, CH-5232 Villigen PSI, Switzerland}}\\[0.5cm]
\end{center}
\begin{abstract}
It is well known that the Mini\-mal Supersymmetric  Standard Model (MSSM) 
can enhance $\br(\Bds)$ by orders of magnitude \cite{beku02}, even if we
assume the Cabib\-bo-Ko\-ba\-ya\-shi-Maskawa (CKM) matrix to be 
the only source of flavour  violation. Of 
particular interest is the quantity $R\equiv\br(\Bd)/\br(\Bs)$, since
(i) the  theoretical errors cancel to a large extent, and (ii) it
offers a  theoretically clean way of extracting the ratio
$|V_{td}/V_{ts}|$  in the Standard Model, which predicts
$R_{\sm}\sim |V_{td}/V_{ts}|^2\sim
O(10^{-2})$. Exploring three different scenarios of modified 
minimal flavour violation ($\overline{\rm MFV}$), 
we find that part of the MSSM parameter space can accommodate  large $\Bds$ 
branching fractions, while being consistent with various experimental
constraints.  More importantly, we show that the ratio $R$ can be as
large as $O(1)$, while the individual branching fractions may be
amenable  to detection by ongoing experiments. We conclude that within 
the MSSM with large $\tan\beta$ the decay rates of $\Bds$ 
be of comparable size even in the case where 
flavour violation is due solely to the CKM matrix.
\end{abstract}

%********************
% END OF TITLE PAGE 
%********************

%***********************
\section{Introduction}
%***********************

The SM predicts the $\Bs$ branching ratio to be \cite{cscp,lectures} 

\be\label{sm:Bs}
 \br(\Bs)=(3.2\pm1.5)\times 10^{-9},
\ee
and the ratio of  branching  fractions
\be\label{rsm}
 R_{\rm SM} \equiv \left. \f{\br(\Bd)}{\br(\Bs)} \right|_{\rm SM}
   \approx \f{\tau_{B_d}}{\tau_{B_s}}\f{M_{B_d}}{M_{B_s}}
   \f{f_{B_d}^2}{f_{B_s}^2} \f{|V_{td}|^2}{|V_{ts}|^2} \sim O(10^{-2}),
\ee
where $\tau_{B_q}$ is the lifetime of the $B_q$ meson, $M_{B_q}$ and
$f_{B_q}$ are the corresponding mass and decay constant.
However, given the SM prediction of $\br(\Bd)\sim O(10^{-10})$,
the $\Bd$ decay is experimentally remote unless it is significantly
enhanced  by new physics.~Thus, the purely leptonic decays of 
neutral $B$ mesons provide an ideal testing ground for physics outside
the  SM, with the current experimental upper bounds \cite{exp:bmumu} 
$\br(\Bd) < 2.8 \times 10^{-7}$ and $\br(\Bs) < 2.0 \times 10^{-6}$,
both given at $\cl{90}$.

The main interest in this talk is in a  qualitative comparison  of
the  $\Bds$ branching fractions in the presence of non-standard
interactions, which can be made by using the ratio
\mathindent150pt
\be\label{ratio:BdBs}
 R\equiv \frac{\br(\Bd)}{\br(\Bs)}.
\ee
\mathindent40pt
Referring to \eq{rsm}, it is important to note that the suppression of
$R$ in the SM  is largely due to the ratio of the
CKM elements.~This dependence
on the CKM factors  allegedly pertains to all models in which the quark 
mixing matrix
is the  only source of flavour violation. It is therefore interesting
to  ask if $R$ could be of the order unity in some non-standard 
models where flavour violation is governed exclusively by the CKM matrix.
Working in the framework of the Minimal
Supersymmetric  Standard Model (MSSM) with a large ratio of Higgs
vacuum  expectation values, $\tan\beta$  (ranging from $40$ to $60$),
we show that such a scenario does exist, and study its consequences for 
the $\Bds$ branching ratios. 

The outline is as follows. ~First, in \sec{mssm}, we 
define modified minimal flavour violation ($\overline{\rm MFV}$) and
discuss briefly three  distinct scenarios within 
the MSSM.~Second, the effective Hamiltonian describing the decays $\Bds$ in
the presence of  non-SM interactions is given 
in \sec{heff:branch}. 
Furthermore, in \sec{num:analysis}, numerical results for the 
branching fractions $\br(\Bds)$
and the  ratio $R$ are presented. Finally,  we summarize and conclude.

%**************************************************************
\section{Modified Minimal Flavour Violation}\label{mssm}
%**************************************************************
%
There exists no unique definition  of minimal flavour violation (MFV)
in the literature 
(see, e.g., \rfs{{oszi},{laplace},{london},{munich},{uut}}). 
The common feature of these MFV definitions is that flavour violation
and/or flavour-changing neutral current (FCNC) 
processes are entirely governed by the 
CKM matrix. On the other hand, they differ, for example, by the
following additional assumptions:
i) there are no new operators present, in addition to those of the SM 
          \cite{lectures, oszi,laplace,{uut}},
ii) FCNC processes are proportional to the same combination of
  CKM elements as in the SM \cite{london} or
iii) flavour transitions occur only in charged currents at tree
  level \cite{{munich}}.
While these ad hoc assumptions are useful for certain considerations,
such as the construction of the universal unitarity triangle \cite{uut}, 
they cannot be justified by symmetry arguments on the level of the 
Lagrangian. For example, the number
of operators with a certain dimension is always fixed by the symmetry of the 
low-energy effective theory.~Whether the
Wilson coefficients are negligible or not, depends crucially  on the
model considered and on the part of the parameter space. Furthermore, the 
requirement that FCNC processes 
are proportional to the same combination of CKM elements as in the SM
fails, for example, in the MSSM and can be retained only after 
further simplifying assumptions.
The last statement in iii) is of pure phenomenological 
relevance in order to avoid huge contributions to FCNC processes.

Using symmetry arguments, we propose an approach that  relies only on the 
key ingredient of the MFV definitions in 
\rfs{{oszi},{laplace},{uut},{london},{munich}}, without considering the 
above mentioned additional assumptions i)--iii). We call an extension of the SM
a modified minimal flavour-violating ($\overline{\rm MFV}$) model if
and only if FCNC processes or flavour violation are
entirely ruled by the CKM matrix; that is, we require that FCNC
processes 
vanish to all orders in perturbation theory in the limit $\vckm \to \openone$.
For a motivation of this definition using symmetry arguments we refer 
the reader to \rf{beku02}.
As will become clear, the advantage of  $\overline{\rm MFV}$ is that it is less
restrictive than MFV, while the CKM matrix remains the only source of 
FCNC transitions.

Our definition of $\overline{\rm MFV}$ is manifest basis independent. 
However, in order to find a useful classification of different
$\overline{\rm MFV}$ scenarios within the MSSM, we will  work
in the super-CKM basis. In this basis
the quark mass matrices are diagonal,
and  both quarks and squarks are rotated simultaneously. The scalar quark 
mass-squared matrices in this basis have the structure 
\be\label{squark:mass}
 \M_U^2 = \left(\begin{array}{cc} \M^2_{U_{LL}} & \M^2_{U_{LR}}\\ 
                \M^{2\dag}_{U_{LR}} & \M^2_{U_{RR}}\end{array}\right), \quad
 \M_D^2 = \left(\begin{array}{cc} \M^2_{D_{LL}} & \M^2_{D_{LR}}\\
                \M^{2\dag}_{D_{LR}} & \M^2_{D_{RR}}\end{array}\right),
\ee
where the $3\times 3$ submatrices are given in \cite{beku02}. Here we present
the two important entries only:
\begin{equation} 
\M^2_{U_{LL}} = \ml{U} + M_U^2  + \frac{1}{6} M_Z^2\cos2\beta(3 - 
4 \sin^2\theta_W)\openone, 
\end{equation}
\begin{equation}          
 \M^2_{D_{LL}} = \ml{D}+ M_D^2 - 
\frac{1}{6}M_Z^2\cos2\beta(3-2\sin^2\theta_W)\openone.
\end{equation}
Because of SU(2) gauge invariance, the mass matrix $ \ml{D}$ is
intimately  connected to $\ml{U}$ via 
\be\label{su2}
 \ml{D}=\vckm^\dag \ml{U} \vckm,
\ee
which is important for our subsequent discussion. 

The above given definition of $\overline{\rm MFV}$ requires that the 
soft SUSY trilinear couplings $A_U$, $A_D$ and the soft SUSY breaking 
squark masses $M_{\tilde U_R}$, $M_{\tilde D_R}$ are diagonal.

Taking into account the relation in \eq{su2}, one encounters three
cases of $\overline{\rm MFV}$. 
\bit
\item {\bf Scenario (A):}\\
$\ml{U}$ is proportional to the unit matrix, and so
$\ml{U}=\ml{D}$.~As a result, there  are no gluino and neutralino
contributions  to flavour-changing transitions at one-loop level. 
This scenario of $\overline{\rm MFV}$ coincides with the
MFV scenario at low $\tan\beta$ as defined in Refs. \cite{{munich},{uut}}.  
\item {\bf Scenario (B):}\\
$\ml{D}$ is diagonal but not proportional to the unit matrix and, in 
consequence,
$\ml{U}$ has  non-diagonal entries.~In such a case, there are again no
gluino and neutralino contributions to flavour-changing one-loop
transitions involving only external down-type quarks and
leptons. However, 
additional chargino contributions show up, due to
non-diagonal entries of $\ml{U}$.
\item {\bf Scenario (C):}\\
$\ml{U}$ is diagonal but not proportional to the unit matrix, which
gives  rise to off-diagonal entries in 
$\ml{D}$.~Accordingly, gluino and neutralino exchange diagrams
(in addition to those involving $W^\pm, \chargino$,
charged  and neutral Higgs bosons)
contribute to flavour-changing transitions at 
one-loop level that involve external down-type quarks.
\eit

The common feature of  all these scenarios is that the CKM matrix is
the only  source of flavour violation.

%************************************************************************
\section{Effective Hamiltonian}\label{heff:branch}
%************************************************************************
The effective Hamiltonian responsible for the processes $\Bq$, with
$q=d,s$  and $l=e,\mu,\tau$, in the presence of non-standard interactions
is  given by
\be\label{heff}
 H_{\rm eff} = -\f{G_F\alpha}{\sqrt{2}\pi} V_{tb}^{}V_{tq}^{*}
               \sum_{i=10,S,P}[c_i(\mu)\op_i(\mu)+c_i'(\mu)\op_i'(\mu)],
\ee
with the short-distance coefficients $c_i^{(\prime)}(\mu)$ and the
local  operators
\be\label{operator:basis}
 {\Oi}_{10}=(\bar{q} \gamma^{\mu} P_L b) (\bar{l} \gamma_{\mu}\gamma_5 l),\quad
 {\Oi}_S= m_b (\bar{q} P_R b) (\bar{l}l),\quad
 {\Oi}_P= m_b (\bar{q} P_R b)(\bar{l} \gamma_5 l),
\ee
where $P_{L,R}= (1\mp \gamma_5)/2$. The primed operators can be obtained by
$P_L \leftrightarrow P_R$ and $m_b\rightarrow m_q$. 
It turns out that the primed Wilson
coefficients are negligibly small, and hence can be safely neglected.

At high $\tan\beta$ the scalar and pseudoscalar operators $\Oi_S$ and $\Oi_P$
become important in addition to the so called SM-operator $\Oi_{10}$. In 
this region of the parameter space an expansion according to $\tan\beta$
is possible and a general expression for $c_i {\cal O}_i$ looks like
\be
c_i {\cal O}_i = \sum_n A_n \tan^{n+1}\beta \;  \left(\frac{m}{M}\right)^n +
                 \sum_n B_n \tan^{n}\beta \;  \left(\frac{m}{M}\right)^n + \ldots,
\ee
where we call the first term leading and the second term subleading.
$m$ denotes lepton and light quark masses while
$M$ stands for masses of particles that have been integrated out.
Explicit expressions for the Wilson-coefficients can be found in
\rfs{{beku02},{cscp},{nlo:susy:RareDecays}}.

%************************************************************
\section{Numerical analysis}\label{num:analysis}
%************************************************************

The experimental bounds used in this numerical analysis as well
as the ranges of the MSSM parameters can be found in Ref. \cite{beku02}.

\subsection*{Scenario (A) and (B)}
Recall that in scenario (A) the matrices $\ml{U}$ and $\ml{D}$ are
equal  and proportional to the unit matrix. Therefore, the gluinos and
neutralinos  do not contribute at one-loop level. The scan over the 
parameter region shows that the ratio $R$
is approximately constant and close to $R_{\rm SM}\approx
0.03$.  

In scenario (B), the matrix $\ml{D}$ is diagonal,
$ \ml{D} = \diag(m^2_{\tilde d_{L}},m^2_{\tilde s_{L}},m^2_{\tilde b_{L}})$,
with at least two different entries. Hence, there are no gluino and
neutralino contributions at one-loop level. Employing the relation in
\eq{su2}, the  matrix $\ml{U}= \vckm  \ml{D} \vckm^\dag$ becomes
non-diagonal. These off-diagonal flavour-changing entries are 
constrained by experimental 
data on $K^0$--$\bar K^0$, $B^0$--$\B^0$,  $D^0$--$\bar D^0$  oscillations, 
and the $b\to s \gamma$ decay \cite{fcnc,bsg:deltas:allcontribs,deltas:bounds}.
It is important to note that the bounds on these flavour-changing 
entries\cite{fcnc} severely constrain the additional chargino 
contributions in scenario (B), hence we end up with a result similar
to scenario (A). $R$ varies between $0.026$ and $0.030$.
Neglecting the constraints we could have found $R$ in the range 
$0.002\lesssim R\lesssim 0.115$. 

\subsection*{Scenario (C)}
In this case, the matrix $\ml{U}$ is diagonal, 
$\ml{U} = \diag(m^2_{\tilde u_{L}},m^2_{\tilde c_{L}},m^2_{\tilde t_{L}})$,
with at least two different entries. According to the relation in
\eq{su2},  this implies that $\ml{D}$ has non-diagonal entries, so
that  gluinos and neutralinos contribute to the $b\to ql^+l^-$
transition already at one-loop level. As before, we take the constraints of 
\rfs{fcnc,bsg:deltas:allcontribs,deltas:bounds} on these off-diagonal 
elements.

%*****************
\begin{figure}[thb]
\begin{center}
\begin{tabular}{cc}
\mbox{\epsfxsize=6.5cm\epsffile{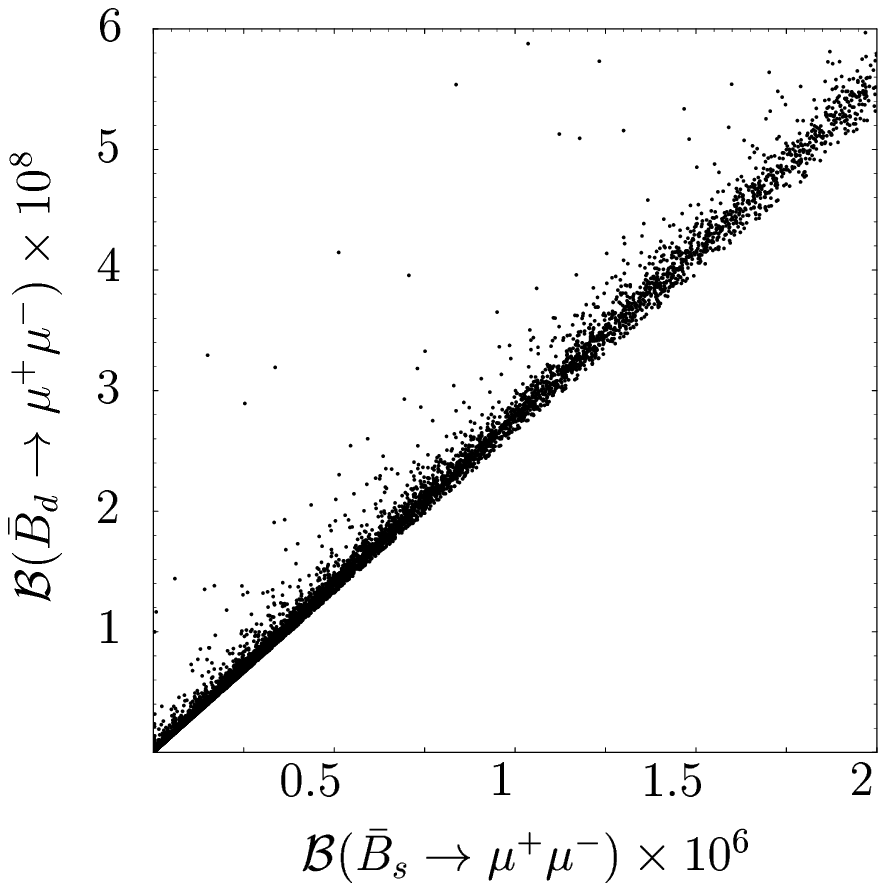}}\hspace{1cm}&
\hspace{1cm}
\mbox{\epsfxsize=6.5cm\epsffile{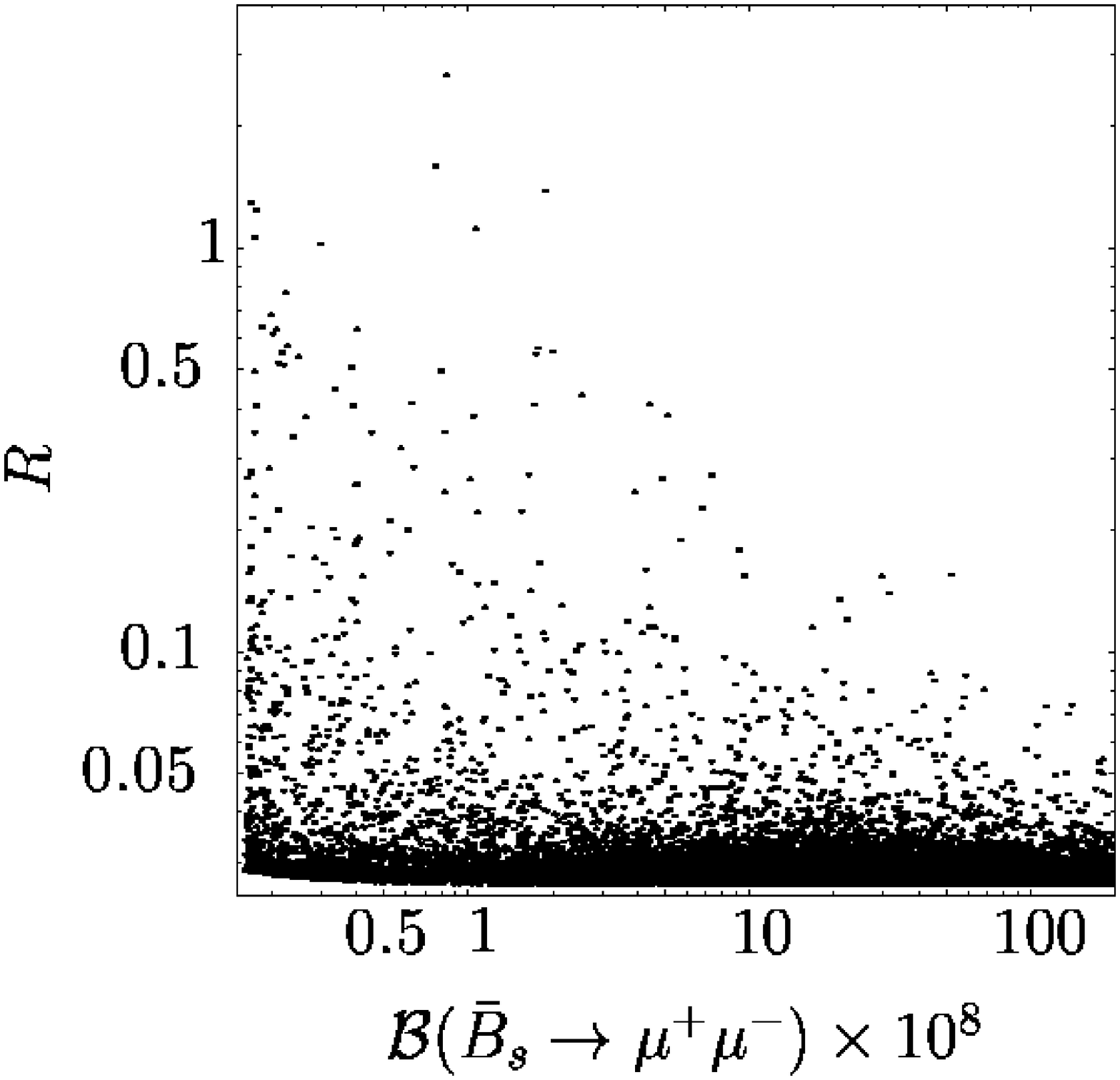}}
\end{tabular}
\vspace{3mm}
\caption{Predictions for the branching ratios $\br(\Bd)$
  vs. $\br(\Bs)$ (left plot) and $R$ vs. $\br(\Bs)$ (right plot) in 
  scenario (C). $R$ varies between $0.026$ and $2.863$.}
\label{scatterscenC}
\end{center}
\end{figure}
%*****************

The scatter plots in Fig.~\ref{scatterscenC} exhibit an order-of-magnitude
deviation from $R_{\rm SM} \approx 0.03$. We find 
$0.026\lesssim R\lesssim 2.863$. A noticeable feature of
scenario (C) is that there exists a lower bound on $R$, i.e. 
$R\gtrsim 0.95\,R_{\rm SM}$ (see left plot of
Fig.~\ref{scatterscenC}), which is due to the structure of the CKM 
matrix. We stress that
this bound is valid only within scenario (C) and does not apply to 
scenario (B) or scenarios with new sources of flavour violation
(see Sec. \ref{mssm}). 

%***********
% FIGURE ?? 
%***********
\begin{figure}[thb]
\begin{center}
\begin{tabular}{cc}
\mbox{\epsfxsize=6.5cm\epsffile{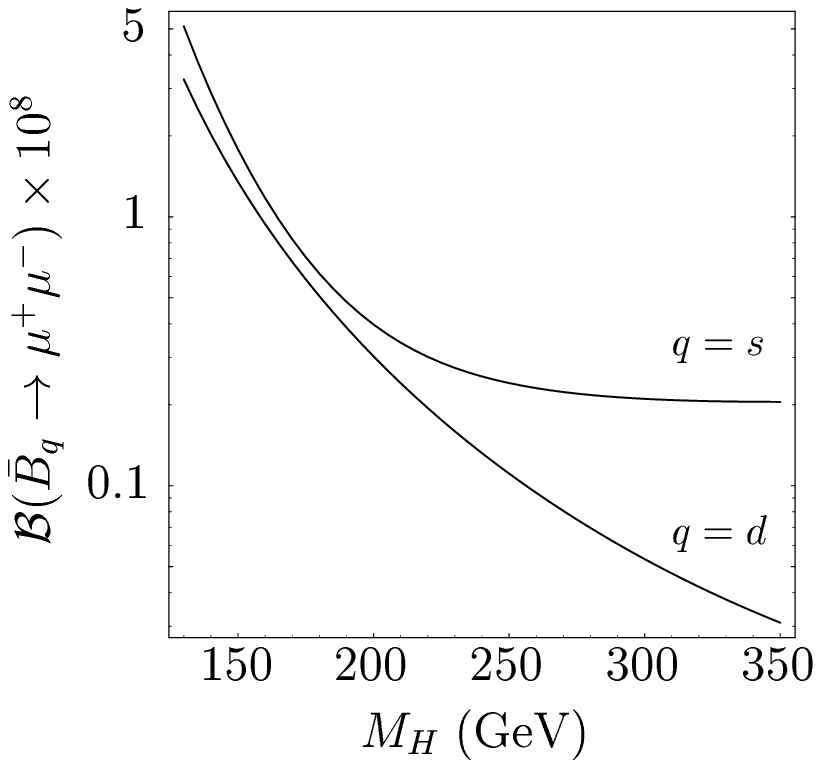}}\hspace{1cm}&
\hspace{1cm}
\mbox{\epsfxsize=6.5cm\epsffile{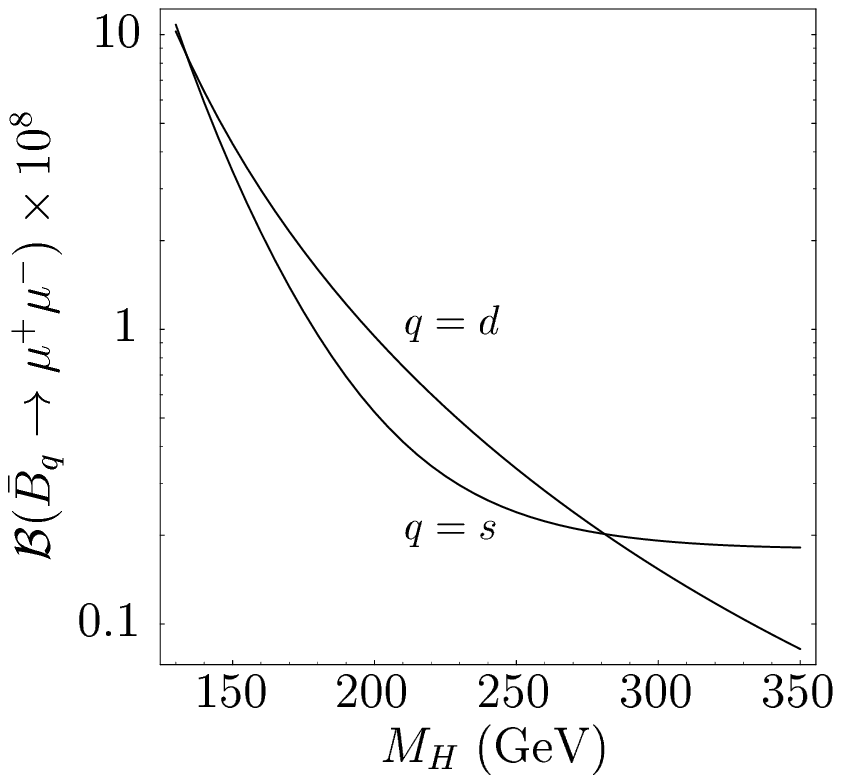}}
\end{tabular}
\vspace{3mm}
\caption{$\br(\Bds)$ as function of the charged Higgs boson mass,
  $M_H$, for $\tan\beta=50$
 (left plot) and $\tan\beta=60$ (right plot), taking into account the 
  experimental constraints on the rare $B$ decays and on the flavour-changing 
  entries. For the remaining parameters see \cite{beku02}.}
\label{mfvc:Bq-mh}
\end{center}
\end{figure}
%***********
% END FIGURE  
%***********

An interesting subset of parameter points was considered in
\fig{mfvc:Bq-mh} (for details see \rf{beku02})
Interestingly, in the left (right) 
plot, the ratio $R$ ranges between $0.15 \lesssim R \lesssim 0.81$ 
($0.44 \lesssim R \lesssim 1.78$), while the magnitude of the individual
branching fractions decreases drastically with increasing charged Higgs 
boson mass, $M_H$. Note that for small $M_H$ and
$\tan\beta$ close to $60$ both branching ratios are in a region that
can be probed experimentally, in Run II of the Fermilab Tevatron,
BABAR and Belle.

The neutralino
Wilson coefficients are numerically smaller than those coming from the
chargino and gluino contributions. However, we have found that in
certain  regions of the MSSM parameter space cancellations
between the chargino and gluino coefficients occur, in which case the
neutralino contributions become important. As a matter of fact, for the
SUSY parameter sets examined, we found that a large
value of $R\equiv\br(\Bd)/\br(\Bs)$ always involves such a
cancellation.

%****************************************************
\section{Summary and conclusions}\label{discussion}
%****************************************************

We have defined $\overline{\rm MFV}$  using symmetry
arguments and have shown that $\overline{\rm MFV}$ is less 
restrictive than MFV, while 
the CKM matrix remains the only source of flavour violation. 
Within the MSSM we have investigated three scenarios that are possible 
within the context of $\overline{\rm MFV}$.  
In particular, we have studied the case where the chargino exchange 
diagrams \cite{cscp} as well as the gluino and neutralino exchange diagrams 
\cite{beku02} contribute besides $W^\pm,H^ \pm,\chargino$ [scenario (C)]. 

Including current experimental data on rare $B$ decays, as well as on
$K, B, D$ meson  mixing, we found that in certain regions of the SUSY
parameter space the branching ratios $\br(\Bd)$ and $\br(\Bs)$ can be
up to the order of $10^{-7}$ and $10^{-6}$ respectively. 
Specifically, we showed that there exist regions in
which the branching fractions of both decay modes are comparable in
size, and may well be accessible to Run II of the Fermilab Tevatron
as well as $R$ can deviate from $R_{\rm SM}$ by orders of magnitude.

We wish to stress that a measurement of the branching ratios $\br(\Bds)$, 
or equivalently,  a ratio $R$ of $O(1)$, does not necessarily imply
the  existence of new flavour violation outside the CKM matrix.
Nevertheless, any observation of these decay modes in ongoing and forthcoming 
experiments would be an unambiguous signal of new physics.

%*****************
\section*{Acknowledgments}
%*****************
I would like to thank Christoph Bobeth, Andrzej J.~Buras, Thorsten Ewerth,
Frank Kr\"uger and Michael Spira for  useful discussions and comments on 
the manuscript. 
This work was supported in part by the German `Bundesministerium 
f\"ur Bildung und Forschung' under contract 05HT1WOA3 and by the 
`Deutsche Forschungsgemeinschaft' (DFG) under contract \mbox{Bu.706/1-1.}

%*************
% REFERENCES 
%*************

\end{document}